\documentclass[%
  twocolumn,
  floatfix,
  pra,%
  superscriptaddress,%
  a4paper%
]{revtex4}

\usepackage[T1]{fontenc}
\usepackage[latin1]{inputenc}
\usepackage[english]{babel}

\usepackage{textcomp}
\usepackage{amsmath}
\usepackage{amsfonts}
\usepackage{amssymb}

\usepackage{graphicx}
\usepackage{color}

\usepackage{userdef_macros}

\begin{document}

\title{Optical Trapping of an Ion}

\author{Ch.\ Schneider}
\author{M. Enderlein}
\author{T. Huber}
\author{T. Schaetz}
\email{tobias.schaetz@mpq.mpg.de}
\affiliation{%
  Max-Planck-Institut f\"{u}r Quantenoptik,
  Hans-Kopfermann-Stra\ss{}e 1,
  D-85748 Garching,
  Germany%
}

\begin{abstract}
  For several decades, ions have been trapped by radio frequency (RF) and
  neutral particles by optical fields.
  We implement the experimental proof-of-principle for trapping an ion
  in an optical dipole trap.
  While loading, initialization and final detection are performed in a RF trap,
  in between, this RF trap is completely disabled and substituted by the optical
  trap.
  The measured lifetime of milliseconds allows for hundreds of oscillations
  within the optical potential.
  It is mainly limited by heating due to photon scattering.
  In future experiments the lifetime may be increased by further detuning the
  laser and cooling the ion.
  We demonstrate the prerequisite to merge both trapping techniques in hybrid
  setups to the point of trapping ions and atoms in the same optical potential.
\end{abstract}

\maketitle


\section{Introduction}

Scientists in the multifaceted fields working with trapped particles
like to cite a 1952 statement by Erwin Schr\"{o}dinger, one of
the founders of quantum mechanics \cite{schroedinger:dino}:
``[\ldots]\ it is fair to state that we are not experimenting with single
particles, any more than we can raise Ichthyosauria in the zoo.''
One year later the first quadrupole mass filter was realized
\cite{paul:massfilter, paul:lecture} with the first single ion in a
radio frequency (RF) trap reported in 1980 \cite{neuhauser:single:ion}.

The physical concepts for trapping neutral and charged particles are
closely related.
Electromagnetic multipole fields act on the charges and dipole moments
(e.g., charge distribution), respectively, to provide confining forces on
the particles in ponderomotive potentials.
The history of trapping neutral particles in dipole traps, however, starts a
decade later than ion trapping.
The optical dipole force was first considered to provide a confining mechanism
in 1962 \cite{askaryan:opt:conf,askaryan:opt:conf:trans}, while the possibility
to trap atoms was first proposed by Letokhov \cite{letokhov:standing:wave}.
Subsequently, the theoretical background of dipole forces was developed
\cite{gordon:dipole:trap,dalibard:dressed:atom}, and in 1986 the first optical
trapping of neutral atoms was reported by Chu et al. \cite{chu:dipole:trap}.
The first single atoms confined in an optical dipole trap were reported in
1999 \cite{ye:single:atom}, again delayed by two decades compared to the first
single trapped ion.

One explanation for the delay of trapping neutral particles is that ion traps
provide potential depths of the order of several $\unit{eV} \approx
\kB \times \unit[10^4]{K}$.
Optical traps in contrast typically store neutral particles up to
$\unit[10^{-3}]{K}$ only.
This discrepancy is mainly due to the comparably large Coulomb forces that can
be exerted on charged particles.
However, the sensitivity to stray electric fields accounts for the paradigm
that ions would not be trappable in optical traps.
We demonstrate that optical trapping of ions is feasible even in the close
vicinity of electrodes and we are able to confine the ion in the dipole
potential for hundreds of oscillation periods.

\section{Experimental Setup}

\begin{figure}[!htb]
  \centering
  \includegraphics[width=0.7\hsize,keepaspectratio]{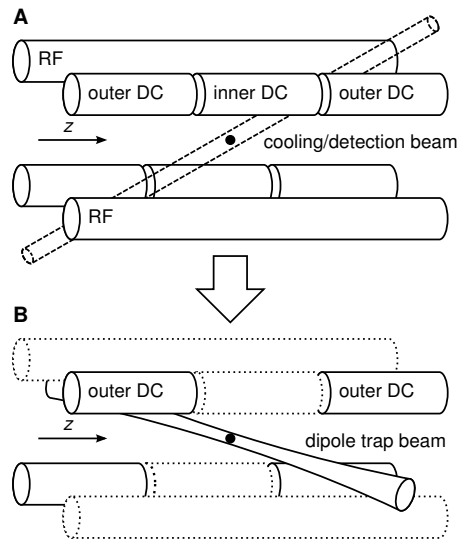}
  \caption{Idealized schemes of the two alternative trapping setups.
    (A) Initialization of the ion while trapping with the
    ponderomotive potential of the RF trap and cooling with the Doppler laser.
    The RF electrodes provide the radial confinement while voltages on the
    outer DC electrodes prevent the ion from escaping axially.
    The arrow labeled ``$z$'' indicates the axis of the RF trap.
    (B) Trapping the ion with the dipole trap laser which crosses the
    axis of the RF trap at an angle of $\unit[45]{\degree}$.
    The ponderomotive potential of the RF trap and the Doppler cooling lasers
    are turned off.}
  \figlabel{setup}
\end{figure}

A single $\atom[24]{Mg}[+]$ ion is created by photo-ionization from neutral
$\atom{Mg}$ atoms evaporated from an oven.
It is Doppler cooled to a few $\unit{mK}$ (Doppler cooling limit:
$\unit[1]{mK}$) in a segmented linear RF trap \cite{schaetz:qsim}
(\figref{setup}A).
The ion is loaded into an optical trap by first superimposing the dipole
potential provided by a Gaussian laser beam tightly focussed on the ion.
Subsequently, the ponderomotive potential of the RF trap is switched off
completely by carefully ramping down its RF drive to zero.
The dipole potential is kept on for the dipole trap duration $T_\text{dipole}$,
while the ponderomotive potential of the RF trap is off
(\figref{setup}B).
Afterwards, the ion is transferred back into the RF trap and detected by
observing its fluorescence light, if the ion remained trapped in the dipole
potential for $T_\text{dipole}$.
During the entire experiment an additional DC field that is focussing in one
dimension is retained to prevent the ion from leaving the dipole trap along the
propagation direction of the beam.
It is crucial to minimize stray electric fields at the position of the ion
and to precisely control the ramp-down of the RF potential.

The dipole trap laser provides a maximal power of around $\unit[500]{mW}$ at a
wavelength of $\unit[280]{nm}$.
The ultra-violet (UV) beam is generated from a $\unit[2]{W}$ fiber laser at
$\unit[1118]{nm}$ by two consecutive second harmonic generation (SHG) stages
\cite{friedenauer:shg}.
Currently, we have up to $P_\text{trap} = \unit[275]{mW}$ at our disposal for
the optical dipole trap.
The efficiency is mainly limited by the losses in an acousto-optical modulator
used for switching the beam, polarization optics providing a
$\sigma^+$-polarization with purity exceeding $1000:1$, and three
telescopes for expanding and cleaning the mode of the beam.
The beam has a nearly Gaussian shape with a waist radius of $w_0
= \unit[7]{\micro m}$.
The frequency of the laser can be detuned up to
$\Delta = - 2 \pi \times \unit[300]{GHz}$ red of the resonance of the
$\level{S}[1/2] \leftrightarrow \level{P}[3/2]$ transition of
$\atom[24]{Mg}[+]$ with a natural line width of
$\Gamma = 2 \pi \times \unit[41.8]{MHz}$ \cite{ansbacher:mg:lifetime}.
According to these parameters we anticipate a depth of the optical potential
of $U_0 \approx \kB \times \unit[51]{mK}$ and a maximum force
perpendicular to the beam of $F_\text{rad} = 2 \times \unit[10^{-19}]{N}$.
The corresponding trapping frequencies at the approximately harmonic bottom of
the trap amount to $\omega_\text{rad} \approx 2 \pi \times \unit[192]{kHz}$
perpendicular and $\omega_\text{k} \approx 2 \pi \times \unit[2]{kHz}$ in the
direction of the beam propagation.
The beam crosses the axis of the RF trap at an angle of $\unit[45]{\degree}$.

The radial frequencies of the RF trap result from the ponderomotive potential
and are initially set to $\omega_{x,y} \approx 2 \pi \times \unit[900]{kHz}$.
The radial trap depth can be estimated to $U_\text{x,y} \approx \kB \times
\unit[10\,000]{K}$ from the minimal electrode--ion distance of
$R_0 = \unit[0.8]{mm}$ and geometrical considerations.
The high voltage for the RF electrodes at a frequency of $\Omega_\text{RF}
\approx 2 \pi \times \unit[56]{MHz}$ is generated by a helical resonator
($1 / e$ ring-down time $T_\text{rd} = \unit[0.5]{\micro s}$).
Initially, the axial frequency is tuned to $\omega_z =
2 \pi \times \unit[115]{kHz}$ by applying appropriate voltages to the outer DC
electrodes (\figref{setup}A).
The axial frequency is lowered to a value of $\omega_z \approx 2 \pi \times
\unit[45]{kHz}$ only shortly before each of the next procedures,
because the ions are more prone to loss by collisions with residual gas at the
corresponding lower axial well depth ($\sim \unit[0.5]{K}$, see methods).

Before the RF confinement can be substituted by the optical dipole trap,
residual static electric fields must be minimized at the position of the ion
to reduce their resulting forces to a level smaller than the maximal force
of the dipole trap.
In our setup there are two main sources of such fields:
Dielectrics in proximity of the trap are charged by the photoelectric
effect due to stray laser light.
Furthermore, each loading process contaminates the trap electrodes which leads
to contact potentials ($\unit[-1.44]{V}$ for $\atom{Mg}$ on $\atom{Au}$).
The minimization of these fields is achieved by gradually ramping down the RF
voltage to approximately $\unit[15]{\%}$ of the initial amplitude and
counteracting the ion's displacement by appropriate voltages applied to all of
the DC segments (see \figref{setup}) and an additional wire placed below the
RF trap.
The voltage applied to an inner DC electrode is fine-tuned on the order of few
$\unit[100]{\micro V}$ corresponding to a residual force in the order of
$\unit[10^{-20}]{N}$ at the position of the ion.

We want to emphasize that a stable three-dimensional confinement by exclusively
DC voltages is impossible according to Earnshaw's theorem:
A DC confinement in one or two dimensions has a defocussing effect in at least
one of the remaining dimensions.
Stable three-dimensional confinement is possible for RF and DC voltages
meeting certain stability criteria (stability diagram \cite{paul:lecture}).
This means that switching off the RF trap by only decreasing the RF voltage
is equivalent to running a scan in a RF mass filter and will result in the
loss of any ion even before a zero RF voltage is reached.
Additionally, there are higher order resonances within the theoretically
stable regime because of anharmonicities of the RF field
\cite{drakoudis:paul:trap:instabilities}.
These effects also have to be considered for the successful transfer of the ion
from the ponderomotive potential of the RF trap to the optical dipole trap.
Therefore, switching off the RF voltage should be performed fast to avoid
excessive heating.
The lower limit is imposed by the ring-down of the helical resonator.
Currently, $\unit[50]{\micro s}$ are chosen to turn off the resonator
sufficiently adiabatic.

After the completed initialization, the trapping attempt is performed according
to the following protocol:

\begin{myitemize}
  \itemsep=0pt%
  \item switching off the Doppler cooling laser
  \item switching on the dipole trap laser
  \item ramping down the ponderomotive potential of the RF trap to effectively
    zero
  \item waiting for the dipole trap duration $T_\text{dipole}$
  \item ramping up the ponderomotive potential of the RF trap
  \item switching off the dipole trap laser
  \item switching on the Doppler cooling lasers
\end{myitemize}

If the trapping attempt is successful, the ion can be detected by observing
its fluorescence light on a CCD camera.
The discrimination between successful and unsuccessful trapping attempts has
an efficiency of effectively $\unit[100]{\%}$.
Trapping attempts according to the above protocol are repeated until the ion
is lost.

\section{Experimental Results}

\begin{figure}[!htb]
  \centering
  \includegraphics[width=0.8\hsize,keepaspectratio]{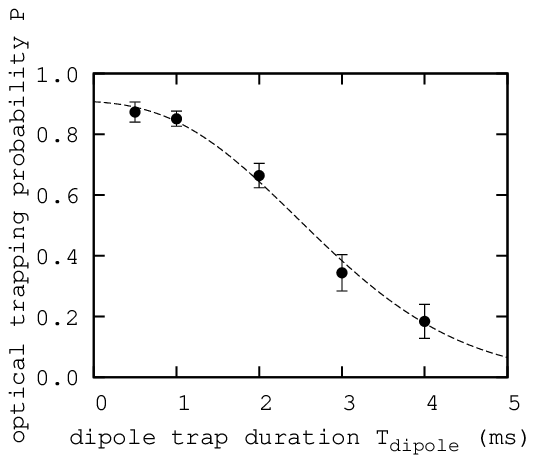}
  \caption{Optical trapping probability $P(T_\text{dipole})$ as a function of
    the trapping duration $T_\text{dipole}$:
    The optical trapping probability is determined as the ratio of successful
    trapping attempts to the total number of trapping attempts.
    Each data point represents experiments on $\sim 40$ ions until each one
    is lost with statistical errors (except for the point at
    $\unit[500]{\micro s}$ based on $17$ ions).
    The fitted curve (dashed line) is based on a simplistic Markovian heating
    model (see methods).
    Experimental parameters: beam waist radius $w_0 = \unit[7]{\micro m}$,
    laser detuning $\Delta = - 2 \pi \times \unit[275]{GHz}$, laser power
    $P_\text{trap} = \unit[190]{mW}$, axial DC frequency
    $\omega_z = 2 \pi \times \unit[47]{kHz}$, RF switched off}
  \figlabel{lifetime}
\end{figure}

In \figref{lifetime} the probability $P(T_\text{dipole})$ for an ion to remain
in the dipole trap is shown as a function of the trapping duration
$T_\text{dipole}$.
Each data point is determined as the number of successful trapping attempts
divided by the total number of attempts with approximately $40$ ions
(corresponding, e.g., to more than $200$ attempts for $T_\text{dipole} =
\unit[1]{ms}$).
For each ion stray fields were minimized as described above.
From the data we calculate a lifetime of $\tau = \unit[(1.8 \pm 0.3)]{ms}$.

Without cooling, the lifetime in an optical dipole trap is limited by heating.
For our experimental parameters with a detuning of $\Delta \approx -6500 \Gamma$
recoil heating due to scattering of laser photons
\cite{gordon:dipole:trap,dalibard:dressed:atom} poses an upper bound on the
lifetime.
The experimental parameters yield a Raman scattering rate of $\Gamma_\text{s}
\approx \unit[860]{ms^{-1}}$ transferring a mean energy of $2 \times E_\text{r}
\approx \unit[10]{\micro K}$ in a single scattering event.
With a dipole trap depth of $U_0 \approx \kB \times \unit[38]{mK}$
($\omega_\text{rad} \approx 2 \pi \times \unit[165]{kHz}$) for the parameters
of \figref{lifetime} this leads to a lifetime of $\tau_\text{theo} \approx
\unit[4]{ms}$.
This estimate assumes a perfect Gaussian beam, perfect $\sigma^+$-polarization,
and zero initial temperature.
Furthermore, it does not assume any heating due to the transfer from the RF
trap into the dipole trap, due to (fluctuating) residual electric fields, or
due to beam pointing instabilities \cite{grimm:dipole:trap}.
Since the measured lifetime is not much less than the estimate, we conclude
that it is mainly limited by photon scattering.

\begin{figure}[!htb]
  \centering
  \includegraphics[width=0.8\hsize,keepaspectratio]{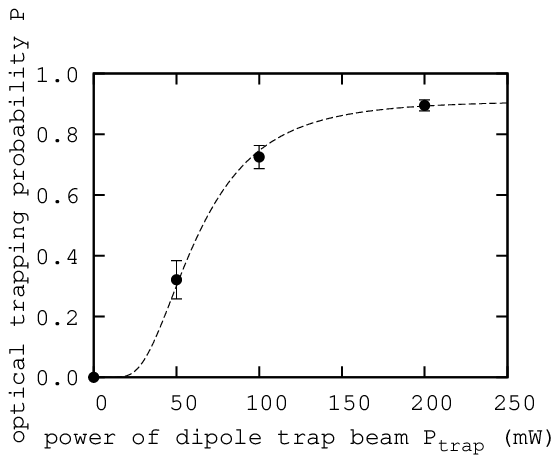}
  \caption{Optical trapping probability $P(P_\text{trap})$ as a function of
    the power of the dipole trap beam $P_\text{trap}$:
    The probability is determined as for \figref{lifetime}, the data points
    again consider statistical errors only, and the fitted curve (dashed line)
    is based on the same simplistic Markovian heating model with identical
    values for the fit parameters (see methods).
    Experimental parameters: beam waist radius $w_0 = \unit[6.4]{\micro m}$,
    laser detuning $\Delta = - 2 \pi \times \unit[300]{GHz}$, dipole trap
    duration $T_\text{dipole} = \unit[500]{\micro s}$, axial DC frequency
    $\omega_z = 2 \pi \times \unit[41]{kHz}$, RF switched off}
  \figlabel{powerscan}
\end{figure}

In \figref{powerscan} the optical trapping probability $P(P_\text{trap})$  is
shown as a function of the beam power $P_\text{trap}$.
The rest of the experimental parameters remains comparable to those in
\figref{lifetime}.

We observe that the trapping probability decreases for decreasing
$P_\text{trap}$ (i.e., depth of the dipole potential).
Considering only recoil heating and zero initial temperature, the lifetime and
thus the optical trapping probability would not depend on the laser power for
$P_\text{trap} > 0$ in our experiment, because the potential depth as well as
the heating rate in the optical dipole trap both scale linearly with
$P_\text{trap}$.
To explain our result we have to take into account, e.g., the non-zero initial
temperature and an increasing sensitivity to the quality of the minimization
of residual electric fields at lower $P_\text{trap}$.
The data point at zero power of the dipole trap beam in \figref{powerscan}
shows that trapping was not achieved without the dipole trap.
This emphasizes that there was no relevant residual trapping potential in the
experiment.

In both Figs.~\ref{fig:lifetime} and \ref{fig:powerscan} the experimental
trapping probability reaches a maximum of $P \approx 0.9$.
Trapping attempts with a duration $T_\text{dipole} \ll \unit[500]{\micro s}$
did not exceed this value significantly either.
Besides a non-zero initial temperature, there are two additional reasons for
that:
Firstly, it is due to the fact that the minimum of the dipole potential did
not always sufficiently overlap with the position of the ion at turn-off of
the RF trap.
The beam path from the last SHG stage to the vacuum recipient currently amounts
to several meters and the collimated beam is focused by a lens with a focal
length of $\unit[200]{mm}$ onto the ion.
This makes the dipole trap beam prone to thermal drifts and shaking caused by
convection of the air.
Moreover, insufficient minimization of residual static electric fields can give
rise to ion loss even at zero $T_\text{dipole}$.

\section{Outlook}

Our experimental setup leaves room for substantial technical improvements
opening up a variety of applications.
The main technical challenges are the limited lifetime and coherences of the
electronic states.
The reduction of the Raman scattering rate $\Gamma_\text{s}$
($\propto P_\text{trap} / \Delta^2$) can be achieved by increasing the detuning
$\Delta$.
Optical ion trapping might still be feasible even at the reduced potential
depth $U_0$ ($\propto P_\text{trap} / \Delta$), if we increased the
electrode--ion distance $R_0$ and therefore reduced the influence of electric
potentials and optical stray fields.
Yet further improvement is anticipated for larger intensity (and $\Delta$)
within an optical cavity \cite{guthohrlein:ion:probe} or by larger
$P_\text{trap}$ encouraged by recent laser development that pioneered
Raman-amplified systems providing $\unit[150]{W}$ of laser power at our
fundamental wavelength of $\unit[1120]{nm}$ \cite{feng:150w:laser} or
$\unit[25]{W}$ at $\unit[589]{nm}$ \cite{feng:25w:laser}, which should be
tunable down to $\unit[560]{nm}$ and allow for efficient SHG.
Depending on the intended experiment, using different ion species (e.g.,
$\atom{Sr}[+]$ or $\atom{Ba}[+]$) would permit wavelengths in the visible
range.
Conventional cooling (Doppler, sideband, Sisyphos cooling) will
further enhance the lifetime and cavity cooling, as demonstrated in
\refcite{leibrandt:cavity:cooling}, might not have to affect the electronic
coherence.

Novel applications have been proposed for hybrid setups, combining
dipole and ponderomotive RF trap potentials
\cite{cirac:microtraps, schmied:ion:optical:lattice}:
Superimposing a commensurate optical lattice on the ordered structure of an
array of RF traps for individual ions or on an ion crystal frozen within one
single RF trap should allow for quantum simulation (QS) experiments
of mesoscopic two- or even three-dimensional quantum systems.
Dependent on the requirements, the different potentials could also
provide confinement in different dimensions.
For example, the RF field could provide the radial confinement for a linear
chain of ions while a blue detuned optical standing wave could realize the
axial confinement within its ``dark'' lattice sites.

Recently, the combination of a magneto-optical and a RF trap was used to
explore cold collisions of atoms and ions \cite{grier:collisions}, however the
collision energy was limited by RF-driven micromotion.
Conventionally trapped ions were proposed as the ultimate ``objective'' of a
microscope \cite{kollath:microscope} or ``read/write head''
\cite{cirac:microtraps}.
For these and most probably further applications it might be advantageous to
exploit the charge of trapped particles, but to avoid micromotion.
Furthermore, if an ion was loaded into the dipole trap differently, e.g., by
photo-ionizing an optically trapped atom, one could avoid electrodes in the
vicinity of the ion.
Still, electrodes for minimizing residual static electric fields will most
probably be required.

Finally, two- and three-dimensional optical lattices might provide an
alternative approach for ion trap architectures \cite{cirac:microtraps,
seidelin:surface-trap, schaetz:qsim, schmied:optimal}.
One can consider, for example, to implement QS with trapped ions
\cite{porras:eff-spin, friedenauer:qmagnet} in larger, higher-dimensional
systems with the Coulomb force providing long range interaction beyond nearest
neighbors.
The optical lattice sites would be regularly but sparsely occupied allowing for
individual addressability and combining lattices of atoms and ions.

To summarize, we achieved trapping an ion optically even in closest vicinity of
electrodes.
Additionally, we have shown that hybrid setups, combining optical and RF
potentials, are capable to cover the full spectrum of composite
confinements---from RF to optical.
The lifetime of the ion in the optical dipole trap is limited by
photon scattering and is thus expected to be improvable by state of the art
techniques.


This work was supported by MPQ, MPG, DFG (SCHA 973/1-4), SCALA and
the DFG Cluster of Excellence ``Munich Centre for Advanced Photonics''.
We thank Hector Schmitz and Robert Matjeschk for preliminary work and
Karim Murr, Roman ``Ramon'' Schmied, and Stephan D\"{u}rr for helpful
discussions.
We also thank Dietrich Leibfried for comments and suggestions on our
manuscript and Ignacio Cirac and Gerhard Rempe for their great intellectual and
financial support.

\section*{Methods: Axial DC Frequency}

The loading process contaminates the trap electrodes with $\atom{Mg}$ from the
oven leading to contact potentials ($\unit[-1.44]{V}$ for $\atom{Mg}$ on
$\atom{Au}$).
A more detailed examination yields an asymmetric axial potential landscape at
a frequency of $\omega_z = 2 \pi \times \unit[38]{kHz}$ that has
a maximum in a distance of approximately $\unit[100]{\micro m}$ from the
position of the ion with a well depth on the order of $\unit[0.5]{K}$.
In comparison, the outer DC electrodes have a distance of $\unit[1.5]{mm}$ to
the ion.
Beyond that maximum the DC fields have a defocussing effect even in axial
direction and the ion is pulled out of the trap.
Without any contact potentials one would expect a focussing effect of the
outer DC electrodes along the $z$-axis over the complete range of
$\pm \unit[1.5]{mm}$ and a well depth which is approximately three orders of
magnitude larger.

\section*{Methods: Simplistic Markovian Heating Model}

The simplistic Markovian heating model assumes a one-dimensional harmonic
potential of finite depth $U_0 \propto P_\text{trap}$ and a Boltzmann
distribution of particle energies.
The probability for leaving the trapping potential is then on the order of
$\tilde{P}(\beta) \approx \exp(-\beta U_0)$ at each attempt with $U_0 =
\epsilon P_\text{trap}$ and $\beta$ defined below; such attempts occur at a
frequency $\Omega_\text{M}$ (not directly related to the dipole trap
frequency), giving a temperature-dependent decay rate of
$\Gamma_\text{M}(\beta) \approx \Omega_\text{M} \tilde{P}(\beta)$.
The differential equation for the probability of having an ion in the dipole
trap is $\dot{P}(t) = - P(t) \Gamma_\text{M}(\beta(t))$, where we assume the
temperature to increase linearly from an initial value:
$(\kB \beta(t))^{-1} = T_0 + \gamma P_\text{trap} t$.
The solution of this differential equation yields the functional form of the
fits in Figs.~\ref{fig:lifetime} and \ref{fig:powerscan} as the probability
$P(T_\text{dipole})$ for the ion to remain within the trap after a fixed
duration $T_\text{dipole}$.
The parameters $\epsilon / \gamma$, $T_0 / \gamma$, $\Omega_\text{M} /
\sqrt{P_\text{trap}}$, and a pre-factor $C_0$ of the function considering
improper initialization are optimized in the fits while $T_\text{dipole}$
and $P_\text{trap}$ are kept fixed.
The deviations of the relevant parameters in Figs.~\ref{fig:lifetime} and
\ref{fig:powerscan} are within less than $\unit[10]{\%}$.
Therefore we have chosen to perform a joint fit for the data.

\bibliography{dipole_trap}
\bibliographystyle{userdef_bibstyle}

\end{document}